\documentclass[conference]{IEEEtran}
\usepackage{amsmath,cite,bm,amsfonts,amssymb,graphicx,color,mathtools,setspace,comment,multirow,xfrac}
\usepackage{epstopdf}
\usepackage{graphicx}
\usepackage{caption}
\usepackage{subcaption}
\usepackage{color}

\newcommand{\be}{\mathbf{e}}
\newcommand{\bE}{\mathbf{E}}

\newcommand{\bg}{\mathbf{g}}
\newcommand{\bG}{\mathbf{G}}

\newcommand{\bH}{\mathbf{H}}
\newcommand{\bI}{\mathbf{I}}

\newcommand{\bn}{\mathbf{n}}
\newcommand{\bP}{\mathbf{P}}
\newcommand{\bq}{\mathbf{q}}
\newcommand{\bR}{\mathbf{R}}

\newcommand{\bW}{\mathbf{W}}
\newcommand{\bx}{\mathbf{x}}
\newcommand{\bX}{\mathbf{X}}
\newcommand{\by}{\mathbf{y}}

\newcommand{\raisecaption}{\vspace{-0.7cm}}

\begin{document}

\title{Deployment Issues for Massive MIMO Systems}
\author{\IEEEauthorblockA{Callum T. Neil\IEEEauthorrefmark{1},
											Mansoor Shafi\IEEEauthorrefmark{2},
											Peter J. Smith\IEEEauthorrefmark{3},
											Pawel A. Dmochowski\IEEEauthorrefmark{1}
											}
\IEEEauthorblockA{\IEEEauthorrefmark{1}
\small{School of Engineering and Computer Science, Victoria University of Wellington, Wellington, New Zealand}}
\IEEEauthorblockA{\IEEEauthorrefmark{2}
\small{Spark New Zealand, Wellington, New Zealand}}
\IEEEauthorblockA{\IEEEauthorrefmark{3}
\small{Department of Electrical and Computer Engineering, University of Canterbury, Christchurch, New Zealand}}
\IEEEauthorblockA{\small{email:\{pawel.dmochowski,callum.neil\}@ecs.vuw.ac.nz,~p.smith@elec.canterbury.ac.nz,~mansoor.shafi@spark.co.nz}}}

\maketitle

\begin{abstract}
In this paper we examine a number of deployment issues which arise from practical considerations in massive multiple-input-multiple-output (MIMO) systems. We show both spatial correlation and line-of-sight (LOS) introduce an interference component to the system which causes non-orthogonality between user channels. Distributing the antennas into multiple clusters is shown to reduce spatial correlation and improve performance. Furthermore, due to its ability to minimize interference, zero forcing (ZF) precoding performs well in massive MIMO systems compared to matched filter (MF) precoding which suffers large penalties. However, the noise component in the ZF signal-to-noise-ratio (SNR) increases significantly in the case of imperfect transmit channel state information (CSI).
\end{abstract}
\section{Introduction}
\label{sec:Introduction}
%
Global mobile data traffic is expected to increase at a compound annual growth rate of 61$\% $ from 2013-2018 \cite{CISCO}. To accommodate these demands, engineers are proposing new technologies for enhanced network capabilities. A symbiosis of three key technologies: multiple-input-multiple-output (MIMO), millimeter wave (mmWave) and small cell technology, has been proposed \cite{ANDREWS,SWINDLEHURST} in the preliminary research to fifth generation wireless system (5G) standardization.
\par
Massive MIMO, which scales up the number of base station (BS) antennas by at least an order of magnitude, provides significant improvements in spectral efficiency, interference mitigation, data rate and robustness \cite{MARZETTA,RUSEK}, relative to conventional MIMO. Additionally, as the number of antennas grows \cite{SMITH}, user channels start to become mutually orthogonal. As a result, massive MIMO can be implemented with inexpensive, low-powered amplifiers \cite{LARSSON}. However, the deployment of such a large number of antennas introduces many practical limiting factors.
\par
This paper is intended to highlight several deployment issues for downlink (DL) massive MIMO by examining practical scenarios. We consider the effects of spatially correlated antenna topologies, imperfect channel state information (CSI), distributed antenna systems and line-of-sight (LOS) propagation on linear precoding performance in massive MIMO systems. All results are presented in terms of matched filter (MF) expected per-user signal-to-interference-plus-noise-ratio (SINR) and zero forcing (ZF) expected per-user signal-to-noise-ratio (SNR).
\par
The contributions of this paper are as follows:
\begin{itemize}
	\item We explore the impact of different antenna topologies on antenna array spatial correlation and system performance.
	\item We demonstrate that distributing an antenna array into multiple clusters reduces spatial correlation and positively impacts system performance.
	\item We consider a Rician fading channel to model the effects of LOS propagation, which we show to adversely affect massive MIMO system performance.
\end{itemize}
%
\section{System Model}
\label{sec:System_Model}
\subsection{System Description}
\label{subsec:System_Description}
We consider a single-cell massive multi-user (MU)-MIMO DL system with a total of $M$ transmit antennas divided equally among $N$ antenna clusters, jointly serving a total of $K<<M$ single-antenna users. At each antenna cluster, the $\frac{M}{N}$ antennas are assumed to be arranged as $\frac{M}{2N}$ pairs of cross-polarized (x-pol) antennas. We assume time division duplex (TDD) operation with uplink (UL) pilots enabling the transmitter to estimate the DL channel via channel reciprocity. On the DL, the $K$ single antenna terminals collectively receive the $K\times 1$ vector
\begin{equation}
	\by = \sqrt{\rho }\bG^{\textrm{T}}\bx + \bn,
\end{equation}
where $\rho $ is the transmit SNR,  $\bx$ is an $M \times 1$ precoded data vector and $\bn$ is a $K\times 1$ noise vector with independent and identically distributed (i.i.d.) $\mathcal{CN}(0,1)$ entries. The transmit power is normalized, $\mathbb{E}\left[ \| \bx\| ^{2}\right] = 1$, i.e., each antenna transmits at a power of $\frac{\rho }{M}$. The $M\times K$ UL channel matrix, $\bG$, is given by
\begin{equation}
	\bG = \begin{bmatrix}	
		\beta _{1,1}^{\frac{1}{2}}\bR_{\textrm{t}}^{\frac{1}{2}}\bH_{1,1}	&	\ldots 	&	\beta _{1,K}^{\frac{1}{2}}\bR_{\textrm{t}}^{\frac{1}{2}}\bH_{1,K}	\\
       		\vdots    			  		 			& 	\ddots 	& 	\vdots    			   		 			\\
       		\beta _{N,1}^{\frac{1}{2}}\bR_{\textrm{t}}^{\frac{1}{2}}\bH_{N,1} & 	\ldots 	& 	\beta _{N,K}^{\frac{1}{2}}\bR_{\textrm{t}}^{\frac{1}{2}}\bH_{N,K}		
     	\end{bmatrix},
\end{equation}
where $\bH_{n,k}\in \mathbb{C}^{\frac{M}{N}\times 1}$ is the i.i.d. channel vector between the $n$th antenna cluster and the $k$th user, $\beta _{n,k}$ is the link gain coefficient, modeling large-scale effects from antenna cluster $n$ to user $k$, while $\bR_{\textrm{t}}$ is the spatial correlation matrix at each antenna cluster, assumed equal for all antenna clusters.
\subsection{Link Gain Model}
\label{subsec:Link_Gain_Model}
We consider the scenario where users are dropped at random locations within a circular coverage region. Antenna clusters are positioned equidistant on the periphery of the coverage region for $N>1$ while for co-located antenna systems, all antenna elements are located in the centre of the circular coverage region. Path loss is then calculated from a conventional distance based  model with i.i.d. log-normal shadowing: $ALd^{-\gamma }$, where $L$ is i.i.d. log-normal shadowing with standard deviation (SD) $\sigma $, $d$ is the link distance, $\gamma $ is the path-loss exponent and $A$ normalizes the $\beta _{n,k}$ values so that $\textrm{max}\{ \beta _{n,k}, n=1,2,\ldots ,N, k=1,2,\ldots ,K\} =\beta _{\textrm{max}}$ \cite{SMITH}.
\subsection{Spatial Correlation Model}
\label{subsec:Spatial_Correlation_Model}
We consider a cross-polarized (x-pol) antenna configuration, with the spatial correlation matrix modeled via \cite{PAULRAJ}
\begin{equation}
	\bR_{\textrm{t}} = \bX_{\textrm{pol}} \odot \bR,
\end{equation}
where the $M\times M$ matrix $\bR$ is the co-polarized (co-pol) spatial correlation matrix, $\odot $ represents the Hadamard product and $\bX_{\textrm{pol}}$ is the $M\times M$ x-pol matrix given by
\begin{equation}
	\bX_{\textrm{pol}} = \mathbf{1}_{\frac{M}{2}} \otimes \begin{bmatrix}
					1 & \sqrt{\delta } \\
					\sqrt{\delta } & 1 
				\end{bmatrix},
\end{equation}
where $\mathbf{1}_{\frac{M}{2}}$ is a $\frac{M}{2}\times \frac{M}{2}$ matrix of ones, $\delta $ denotes the cross-correlation between the two antenna elements in the x-pol configuration and $\otimes $ represents the Kronecker product.
\subsection{Imperfect CSI Model}
\label{subsec:Imperfect_CSI_Model}
We model CSI imperfections via an \emph{estimated} channel matrix, $\hat{\bG}$, given by \cite{SURAWEERA}
\begin{equation} \label{impCSIG}
	\hat{\bG} = \xi \bG+\sqrt{1-\xi ^{2}}\bE,
\end{equation}
where $\bE$ is independent and statistically identical to $\bG$ and $\xi $, $0\leq \xi \leq 1$, controls the accuracy of the CSI.
\subsection{Linear Precoders}
\label{subsec:Linear_Precoders}
Due to their simplicity and optimality in massive MU-MIMO systems \cite{GAO}, we examine the performance MF and ZF linear precoding techniques.
\subsubsection{MF Precoding}
\label{subsubsec:MF_Precoding}
The MF precoder (also known as a matched beamformer (MBF) and maximum ratio transmission (MRT)) is the most computationally inexpensive precoding technique, allowing the design of many inexpensive antennas ideal for massive MIMO systems. MF precoding in a MU-MIMO system aims at maximizing the received power at each user while neglecting the effects of interference to the other co-scheduled users. As the number of transmit antennas increases without bound, for a fixed number of users and perfect
CSI, MF precoding benefits from the law of large numbers, effectively eliminating all inter-user interference \cite{CLERCKX}. As a consequence, for an infinite number of antennas, all user channels become mutually orthogonal and information capacity is maximized.
\par
For a MF precoder, the $M\times 1$ precoded data vector, with CSI inaccuracies, is given by \cite{SMITH2}
\begin{equation}
	\bx = \frac{1}{\sqrt{\gamma _{\textrm{MF}}}}\hat{\bG}^{\ast }\bq,
\end{equation}
where $\bq $ is the $K\times 1$ data symbol vector, with $\mathbb{E}\left[ ||\bq||^{2} \right] =1$, and 
\begin{equation} 
	\gamma _{\textrm{MF}} = \frac{\textrm{tr}(\hat{\bG}^{\textrm{T}}\hat{\bG}^{\ast })}{K},
\end{equation}
normalizes the average power of the MF precoder. The expected value of the $i$th user's MF SINR is then given by \cite{SMITH2}
\begin{align} \label{mf_sinr}
	&\mathbb{E}\left[ \textrm{SINR}_{i}\right] \notag \\
	&~\approx \frac{
	\frac{\rho }{K\gamma _{\textrm{MF}}}\left( \xi ^{2}\left| \hat{\bg}_{i}^{\textrm{T}}\hat{\bg}_{i}^{\ast }\right| ^{2} +(1-\xi ^{2}) \hat{\bg}_{i}^{\textrm{T}}\bP_{i}\hat{\bg}_{i}^{\ast } \right)
	}
	{
	\frac{\rho }{K\gamma _{\textrm{MF}}}\sum\limits_{k=1,k\neq i}^{K}{\left( \xi ^{2}\left| \hat{\bg}_{i}^{\textrm{T}}\hat{\bg}_{k}^{\ast }\right| ^{2} +(1-\xi ^{2}) \hat{\bg}_{k}^{\textrm{T}}\bP_{i}\hat{\bg}_{k}^{\ast } \right) } + 1
	},
\end{align}
where $\bP_{i} = \mathbb{E}\left[ \be_{i}^{\ast }\be_{i}^{\textrm{T}}\right] $, $\be_{i}$ is the $i$th column of $\bE$, $\hat{\bg}_{i}$ is the $i$th column of $\hat{\bG}$ and the noise power is normalized to 1.
\subsubsection{ZF Precoding}
\label{subsubsec:ZF_Precoding}
The ZF precoding technique forces all intra-cell interference to zero by using coherent superposition of wavefronts to send null vectors to all other users. The ZF precoder is more computationally expensive than the MF precoder as it requires a matrix inverse of $\hat{\bG}^{\textrm{T}}\hat{\bG}^{\ast }$. However with large antenna numbers $\bW = (\hat{\bG}^{\textrm{T}}\hat{\bG}^{\ast })/M$ tends to the identity matrix, $\bI_{K}$, and the computation of the matrix inverse becomes trivial.
\par
For a ZF precoder, the $M\times 1$ precoded data vector, with CSI inaccuracies, is given by \cite{SMITH}
\begin{equation}
	\bx = \frac{1}{\sqrt{\gamma _{\textrm{ZF}}}}\hat{\bG}^{\ast }(\hat{\bG}^{\textrm{T}}\hat{\bG}^{\ast })^{-1}\bq,
\end{equation}
where 
\begin{equation}
	\gamma _{\textrm{ZF}} = \frac{\textrm{tr}((\hat{\bG}^{\textrm{T}}\hat{\bG}^{\ast })^{-1})}{K},
\end{equation}
normalizes the average power of the ZF precoder. The expected value of the $i$th users ZF SNR is then given by
\begin{equation} \label{zf_snr}
	\mathbb{E}\left[ \textrm{SNR}_{i}\right] \approx \frac{\frac{\rho }{K\gamma _{\textrm{ZF}}}\xi ^{2}}{\rho (1-\xi ^{2})\textrm{tr}(\bP_{i})+1}, 
\end{equation}
where the noise power is normalized to 1.
%
\section{Antenna Array Topologies}
\label{sec:Antenna_Array_Topologies}
\begin{figure}
        \centering
        \begin{subfigure}[b]{0.22\textwidth}
        	\centering\includegraphics[trim=2.5cm 9.2cm 3cm 5cm,clip,width=1\columnwidth]{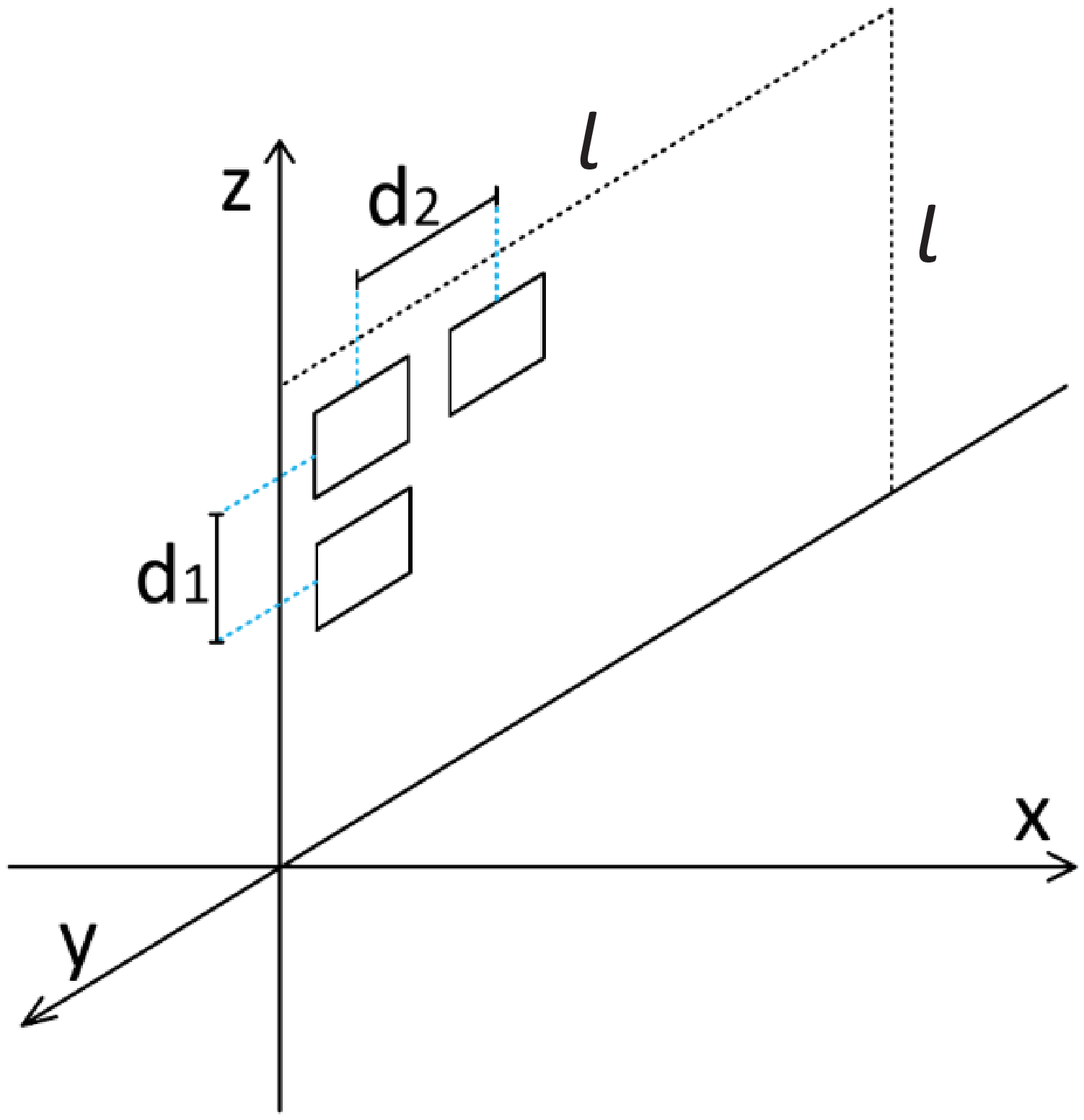}
                \caption{}
                \label{fig:ura}
        \end{subfigure}
        ~
        \begin{subfigure}[b]{0.22\textwidth}
		\centering\includegraphics[trim=2cm 6.5cm 4.2cm 4cm,clip,width=1\columnwidth]{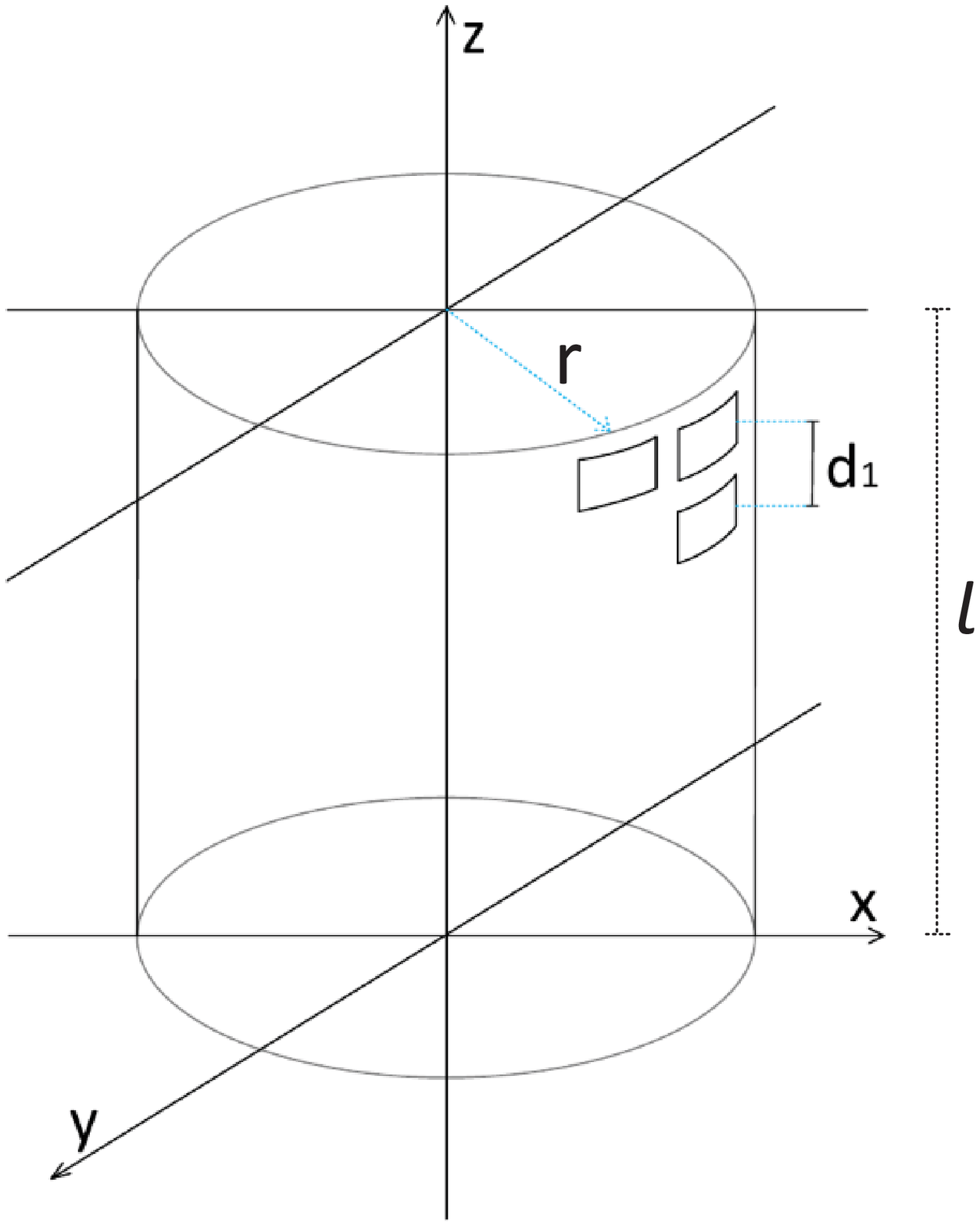}
		\caption{}
                \label{fig:cylindrical}
        \end{subfigure}
	\caption{(a) URA. (b) Cylindrical Array.}
\end{figure}	
It was shown in \cite{SMITH} that the additional benefits expected of massive MIMO systems are realized only when the number of antennas is on the order of a thousand. With such stringent requirements, antenna arrays can become unfeasibly large, e.g., $256$ co-polarized antenna elements positioned at a half-wavelength spacing at $f=2.6$ GHz requires a uniform linear array (ULA) of 14.8 m. To combat space requirements, antenna arrays need to utilize both azimuth and zenith dimensions \cite{NEIL}. We explore the impact on linear precoding performance of deploying antennas in uniform rectangular array (URA) and cylindrical antenna topologies, shown in Figures \ref{fig:ura} and \ref{fig:cylindrical}, respectively. 
\par
We observe the effects of spatial correlation, we constrain the array dimension to $l$ (the width and the height of URA topology). The cylindrical array can then be obtained by wrapping the URA around a virtual cylinder. Thus, in Figure \ref{fig:cylindrical}, the circumference and radius of the uniform circular array (UCA) on the $x,y$-plane is equal to $l$ and $r=2\pi /l$, respectively. Spatial correlation for each antenna topology is then generated via
\begin{equation} \label{R}
	\bR = \mathbf{AF}_{\phi }\mathbf{AF}_{\theta }^{\textrm{H}},
\end{equation}
where $\mathbf{AF}_{\phi }$ and $\mathbf{AF}_{\theta }$ are the azimuth and zenith domain antenna array factors of size $P\times 1$ and $Q\times 1$ respectively, given by \cite{BALANIS}
\begin{equation}
	\mathbf{AF}_{\phi } = \left[ 1, \textrm{e}^{-j\Phi _{1}(\Delta \phi ,\Delta \theta )}, \ldots ,\textrm{e}^{-j\Phi _{P-1}(\Delta \phi ,\Delta \theta )}\right] ^{\textrm{T}},
\end{equation}
\begin{equation}
	\mathbf{AF}_{\theta } = \left[ 1, \textrm{e}^{-j\Theta _{1}(\Delta \theta )}, \ldots ,\textrm{e}^{-j\Theta _{Q-1}(\Delta \theta )}\right] ^{\textrm{T}}.
\end{equation}
$\Phi _{p}(\Delta \phi ,\Delta \theta )$ and $\Theta _{q}(\Delta \theta )$ are the azimuth and zenith domain phase shifts of the $p$th and $q$th antenna's angle of departure (AOD) respectively, with respect to a reference antenna, given by
\begin{equation}
	\Phi _{p}(\Delta \phi ,\Delta \theta ) = kd_{p}\cos (\phi +\Delta \phi )\sin (\theta +\Delta \theta ) ,
\end{equation}
\begin{equation}
	\Theta _{q}(\Delta \theta )=kd_{q}\cos (\theta +\Delta \theta) ,
\end{equation}
where $k$ is the wavenumber, $d_{p}$ is the distance between the $p$th antenna and the reference antenna element in the azimuth domain, $d_{q}$ is the distance between the $q$th antenna and the reference antenna element in the zenith domain, $\Delta \phi $ is the azimuth domain AOD offset relative to its mean $\phi $, and $\Delta \theta $ is the zenith domain AOD offset relative to its mean $\theta $. We model the probability density functions (pdfs) of LOS and non-line-of-sight (NLOS) AODs as described by the 3rd Generation Partnership Project (3GPP) \cite{3GPP} 3D channel model.
\par
All results are generated by simulating MF $\mathbb{E}[\textrm{SINR}_{i}]$ and ZF $\mathbb{E}[\textrm{SNR}_{i}]$, in \eqref{mf_sinr} and \eqref{zf_snr}, respectively, over many channel realizations. System parameters given in Table \ref{table:System_Parameters}, where $\lambda $ is the wavelength.
\par
\begin{table}[ht] 
	\centering 
	\begin{tabular}{c|c} 
		Parameter & Value \\ [0.5ex] 
		\hline\hline 
		Transmit SNR, $\rho $ & 10 dB \\ 
		Log-normal shadowing SD, $\sigma $ & 8 dB \\ 
		Path-loss exponent, $\gamma $ & 4 \\  
		Link distance, $d$  & 50-1000 m \\  
		Maximum link gain coefficient, $\beta _{\textrm{max}}$ & 25 dB \\  
		System frequency, $f$ & 2.6 GHz \\  
		X-pol parameter, $\delta $  & $0.1^{2}$ \\  
		Array dimension, $l$  & $2\lambda $ \\  
		Azimuth domain AOD pdf, $p_{\Delta \phi }(\Delta \phi )$ & Wrapped Gaussian \\
		Zenith domain AOD pdf, $p_{\Delta \theta }(\Delta \theta )$ & Laplacian \\ 
		NLOS mean azimuth AOD, $\phi $ & $74.13^{\circ }$ \\  		
		NLOS mean zenith AOD, $\theta $ & $18.20^{\circ }$ \\  		
		NLOS azimuth AOD SD, $\sigma _{\Delta \phi }$ & $1.29^{\circ }$ \\
		NLOS zenith AOD SD, $\sigma _{\Delta \theta }$ & $1.45^{\circ }$ \\
		LOS mean azimuth AOD, $\phi $ & $64.57^{\circ }$ \\  		
		LOS mean zenith AOD, $\theta $ & $8.91^{\circ }$ \\  	
		LOS azimuth AOD SD, $\sigma _{\Delta \phi }$ & $1.58^{\circ }$ \\	
		LOS zenith AOD SD, $\sigma _{\Delta \theta }$ & $1.45^{\circ }$ \\
		[1ex] 
		\end{tabular} 
		\caption{System Parameters}
		\label{table:System_Parameters} 
\end{table} 
%
\begin{figure}[ht]
	\centering\includegraphics[width=1\columnwidth]{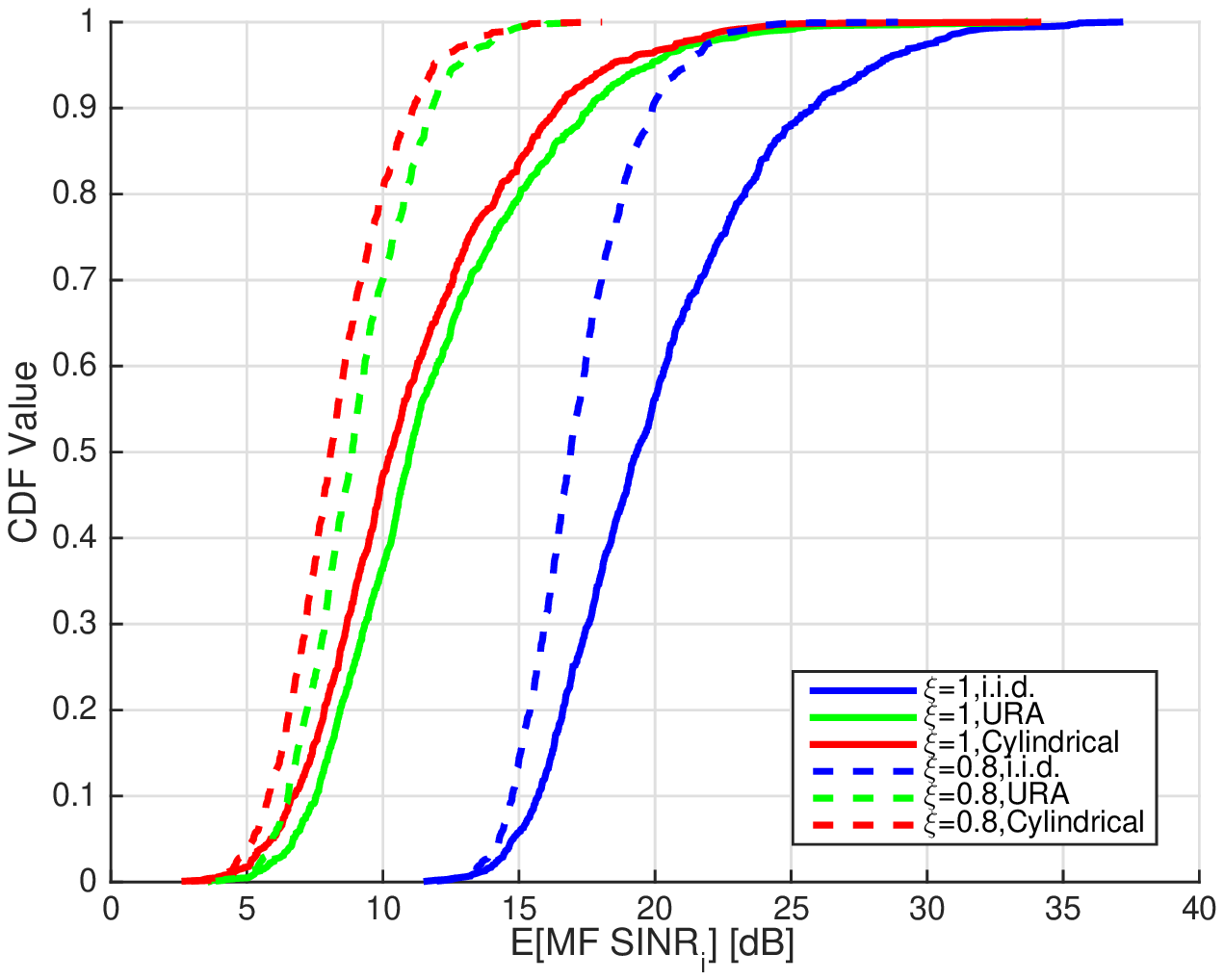}
	\raisecaption\caption{MF $\mathbb{E}[\textrm{SINR}_{i}]$ CDF as a function of CSI accuracy, $\xi $, and antenna topology for $M=256$ and $K=8$.}
	\label{fig:antenna_topology_mf_sinr}
\end{figure}
%
\begin{figure}[ht]
	\centering\includegraphics[width=1\columnwidth]{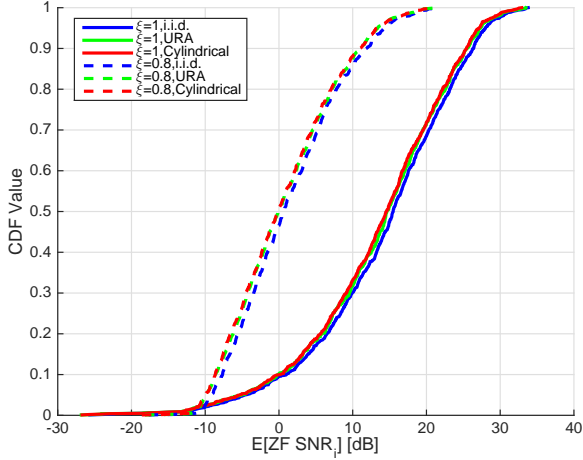}
	\raisecaption\caption{ZF $\mathbb{E}[\textrm{SNR}_{i}]$ CDF as a function of CSI accuracy, $\xi $, and antenna topology for $M=256$ and $K=8$.}
	\label{fig:antenna_topology_zf_snr}
\end{figure}
In Figures \ref{fig:antenna_topology_mf_sinr} and \ref{fig:antenna_topology_zf_snr} we show the impact of URA and cylindrical antenna topologies on the MF and ZF precoding. We plot the cumulative density functions (CDFs) of per user SINR and per user SNR for MF and ZF, respectively. In each case of precoding technique, it can be observed that the URA topology has a marginally better performance than the cylindrical topology. This is a result of smaller inter-element distances in the azimuth domain of the cylindrical array topology compared with the URA. Note, that Maxwell's equations fundamentally limit the effectiveness of 3D arrays such that only antennas on the surface of the array contribute to the information capacity \cite{RUSEK}. The large tails of the ZF precoder CDFs, in Figure \ref{fig:antenna_topology_zf_snr}, are due to the sub-optimality of the precoder at low SNR, where noise becomes dominant. The ZF precoder is more robust to the effects of spatial correlation than the MF precoder, i.e., the reduction in ZF SNR is much less than the reduction in MF SINR when spatial correlation is introduced. This is because spatial correlation acts as a form of interference to the system which the ZF precoder is able to minimize. The MF precoder, on the other hand, is more robust to the effects of imperfect CSI. The additional noise component in the denominator of the ZF SNR, in \eqref{zf_snr}, is increased significantly as $\xi $ is reduced.
\par
%
\begin{figure}[ht]
	\centering\includegraphics[width=1\columnwidth]{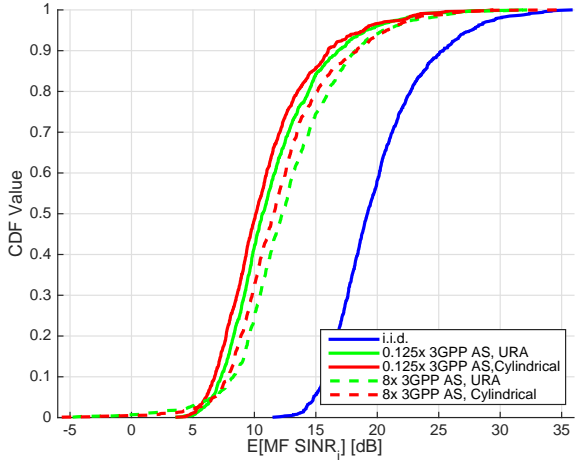}
	\raisecaption\caption{MF $\mathbb{E}[\textrm{SINR}_{i}]$ CDF as a function of angle spread and antenna topology for $M=256$ and $K=8$.}
	\label{fig:as_mf_sinr}
\end{figure}
%
\begin{figure}[ht]
	\centering\includegraphics[width=1\columnwidth]{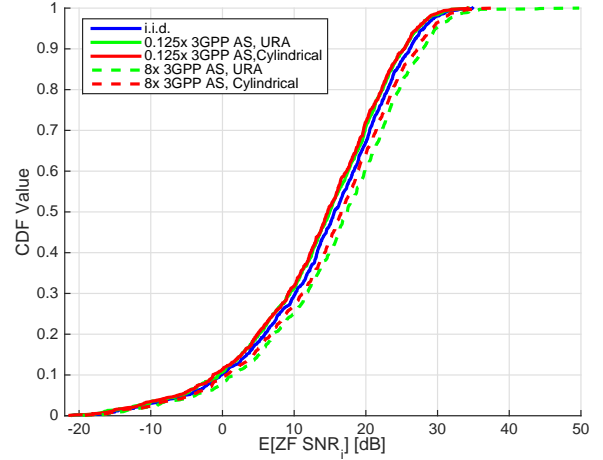}
	\raisecaption\caption{ZF $\mathbb{E}[\textrm{SNR}_{i}]$ CDF as a function of angle spread and antenna topology for $M=256$ and $K=8$.}
	\label{fig:as_zf_snr}
\end{figure}
In Figures \ref{fig:as_mf_sinr} and \ref{fig:as_zf_snr} we present the impact of varying angle spread (AS) on URA and cylindrical antenna topologies. We consider the extreme cases of 0.125x and 8x the 3GPP angle spreads (given in Table \ref{table:System_Parameters}) on MF and ZF precoding techniques. We observe negligible difference in ZF precoding performance for all angle spreads considered. However, the performance of MF precoding is improved slightly with a larger angle spread, e.g., the cylindrical antenna topology shows an increase of median value by approximately 2 dB $\mathbb{E}[\textrm{SINR}_{i}]$. This is a result of a larger angle spread increasing the number of independent paths from each antenna, thus decreasing spatial correlation.
%
\section{Co-located vs Distributed Antenna Systems}
\label{sec:Co-located_vs_Distributed_Antenna_Systems}
\begin{figure}[ht]
	\centering\includegraphics[width=1\columnwidth]{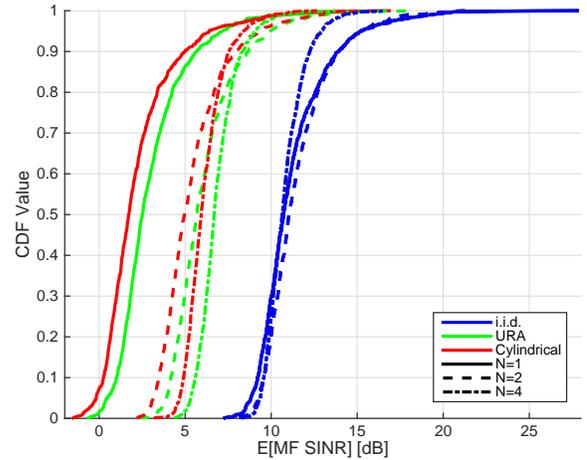}
	\raisecaption\caption{MF $\mathbb{E}[\textrm{SINR}_{i}]$ CDF as a function of antenna cluster numbers, $N$, and antenna topology for $M=256$ and $K=32$.}
	\label{fig:distributed_antennas_mf_sinr}
\end{figure}

\begin{figure}[ht]
	\centering\includegraphics[width=1\columnwidth]{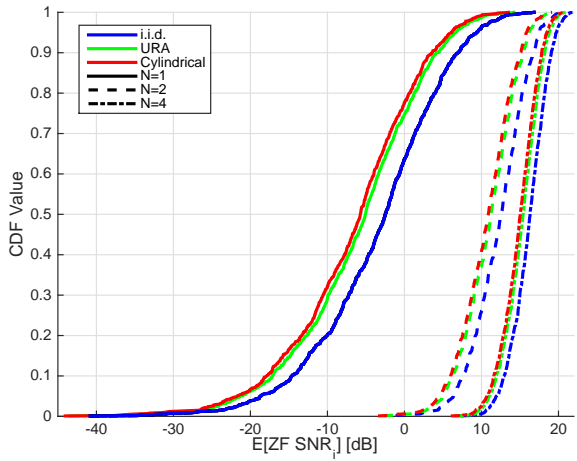}
	\raisecaption\caption{ZF $\mathbb{E}[\textrm{SNR}_{i}]$ CDF as a function of antenna cluster numbers, $N$, and antenna topology for $M=256$ and $K=32$.}
	\label{fig:distributed_antennas_zf_snr}
\end{figure}

The deployment of such a large number of antennas, required for massive MIMO \cite{SMITH} to experience mutual orthogonality between user channels, in confined antenna array dimensions results in significant increases in spatial correlation over conventional MIMO. In order to mitigate spatial correlation, antennas can be distributed into multiple clusters, while assuming equal antenna array form factors. In Figures \ref{fig:distributed_antennas_mf_sinr} and \ref{fig:distributed_antennas_zf_snr} we examine the effects of distributed antennas on linear precoding performance \cite{SMITH2}.
%
%
\par
In Figure \ref{fig:distributed_antennas_mf_sinr} we plot the MF $\mathbb{E}[\textrm{SINR}_{i}]$ CDF for $N=1,2$ and $4$ antenna clusters. For the i.i.d. channel, antenna cluster numbers have a negligible impact, other than at high SNR where the co-located antenna system dominates due to a larger number of degrees of freedom serving users in good coverage. When spatial correlation is introduced, larger numbers of antenna clusters show superior performance. This follows, with equal antenna form factors, as greater numbers of antenna clusters increase inter-element distances and thus reduce the effect of spatial correlation. 
\par
Figure \ref{fig:distributed_antennas_zf_snr} shows the effects of varying antenna cluster numbers, $N$, on ZF $\mathbb{E}[\textrm{SNR}_{i}]$. Contrary to MF precoding, the ZF precoder has significantly better performance with larger antenna cluster numbers for the uncorrelated (i.i.d.) scenarios. This is because $\frac{M}{N}>>K$ for all cases of $N$ considered and so the ZF is able to minimize all system interference coming from the large number of co-scheduled users in the system ($K=32$). The gain in median i.i.d. $\mathbb{E}[\textrm{SNR}_{i}]$ of approximately 20 dB, from $N=1$ to $N=4$, is therefore coming from the better coverage which distributed antennas provides, increasing the link gains, $\beta _{n,k}$. On the other hand, the MF precoders optimality is dependent on the number of degrees of freedom. By distributing the antennas into multiple clusters, we are reducing the number of effective degrees of freedom, e.g., a user close to the co-located antenna system is receiving a strong desired signal from $M=256$ antennas, whereas a user close to the $N=4$ antenna clusters system is only receiving a strong desired signal from $M=64$ antennas. This decrease in effective number of degrees of freedom reduces the ability of the MF precoder to mitigate inter-user interference, which manifests itself as off-diagonal elements in $(\hat{\bG}^{\textrm{T}}\hat{\bG}^{\ast })/M$. Note, that in the limit of an infinite number of degrees of freedom, user channels become orthogonal, i.e., inter-user interference is eliminated, and the the performance of the MF precoder will approach ZF.
%
\section{Massive MIMO Channel Model}
\label{sec:Massive_MIMO_Channel_Model}
With the deployment of small cell technology coexisting with massive MIMO in next generation wireless systems, there is a higher probability of LOS transmission \cite{NEIL2}. The effects of LOS propagation on large antenna array systems has not been well studied and thus we explore its impact on massive MIMO system performance. We model the presence of LOS via a Rician channel with varying K-factor.
\par
The $\frac{M}{N}\times 1$ i.i.d. Rician channel vector $\bH_{n,k}$, between the $n$th antenna cluster and the $k$th user, is given by \cite{XIAO,3GPP}
\begin{equation}
	\bH_{n,k} = \sqrt{\frac{1}{K_{\textrm{f}}+1}}\bH_{n,k\textrm{(NLOS)}} + \sqrt{\frac{K_{\textrm{f}}}{1+K_{\textrm{f}}}}\bH_{n,k\textrm{(LOS)}},
\end{equation}
where the $\frac{M}{N}\times 1$ vector $\bH_{n,k\textrm{(NLOS)}}$, with $\mathcal{CN}(0,1)$ entries, accounts for NLOS propagation and $\bH_{n,k\textrm{(LOS)}}=\mathbf{1}_{\frac{M}{N}\times 1}$ is an $\frac{M}{N}\times 1$ vector of ones which accounts for LOS propagation. $K_{\textrm{f}}$ is the Rician K-factor which controls the dominance of the LOS component relative to the NLOS component.
\par
%
\begin{figure}[ht]
	\centering\includegraphics[width=1\columnwidth]{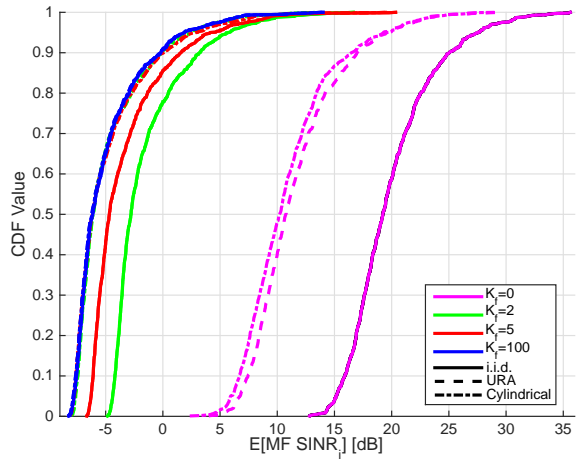}
	\raisecaption\caption{MF $\mathbb{E}[\textrm{SINR}_{i}]$ CDF as a function of Rician K-factor, $K_{\textrm{f}}$, and antenna topology for $M=256$ and $K=8$.}
	\label{fig:channel_model_mf_sinr}
\end{figure}
%
\begin{figure}[ht]
	\centering\includegraphics[width=1\columnwidth]{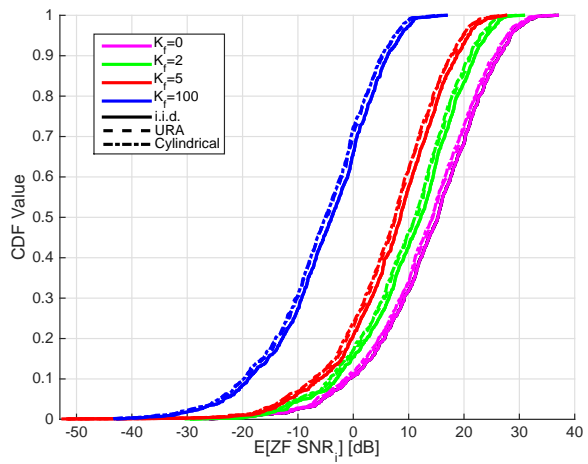}
	\raisecaption\caption{ZF $\mathbb{E}[\textrm{SNR}_{i}]$ CDF as a function of Rician K-factor, $K_{\textrm{f}}$, and antenna topology for $M=256$ and $K=8$.}
	\label{fig:channel_model_zf_snr}
\end{figure}
The impact of varying amounts of LOS propagation on MF $\mathbb{E}[\textrm{SINR}_{i}]$ and ZF $\mathbb{E}[\textrm{SNR}_{i}]$ is shown in Figures \ref{fig:channel_model_mf_sinr} and \ref{fig:channel_model_zf_snr} respectively. For both antenna topologies the $M=256$ transmit antennas are configured as 16 x-pol antennas in the $z$-dimension and 8 x-pol antennas in $x,y$-plane. For the URA, this corresponds to a $\lambda /8$ and $\lambda /4$ adjacent inter-element spacings respectively. The small inter-element spacings produce a large immediate off-diagonal entry in $\bR$, given in \eqref{R}, of nearly 0.94.
\par
It is observed in Figure \ref{fig:channel_model_mf_sinr} that an increased LOS presence, or increasing K-factor, drastically reduces the MF $\mathbb{E}[\textrm{SINR}_{i}]$ performance. For example, increasing the K-factor from $K_{\textrm{f}}=0$ to $K_{\textrm{f}}=2$ decreases the median MF $\mathbb{E}[\textrm{SINR}_{i}]$ of an i.i.d. channel by approximately 22 dB. Even with a very large inter-element spatial correlation factor at 0.94, the effects of any LOS propagation show a much more detrimental effect over spatial correlation for MF $\mathbb{E}[\textrm{SINR}_{i}]$ performance.
\par
As compared with MF precoding, in Figure \ref{fig:channel_model_zf_snr} it can be seen that ZF is more resilient to the effects of LOS propagation. The presence of LOS forms another kind of correlation which the ZF precoder is able to minimize, thus reducing its detrimental effect. However, any LOS propagation conditions produce sub-optimal performance.
%
\section{Conclusion}
\label{sec:Conclusion}
In this paper, we have shown the impact of various practical aspects on massive MIMO system performance. We found that multiple clusters reduces the effects of spatial correlation, and in turn improve linear precoding performance. There is shown to be a negligible difference in system performance between the cylindrical and URA antenna topologies. Also, the ZF precoder is able to minimize the virtual interference introduced by LOS presence and spatial correlation, resulting in superior performance over MF precoding.
%
%
\bibliographystyle{IEEEtran}
\bibliography{bibliography}
\end{document}